\documentclass[prl,aps,twocolumn,showpacs,groupauthors]{revtex4}

\usepackage[english]{babel}
\usepackage[ansinew]{inputenc}
\usepackage{amsmath}                                                           % AMS extensions.
\usepackage{graphicx}                                                          % Include figure files.
\usepackage{bm}                                                                % For bold math.

\newcommand{\ket}[1]{|{#1}\rangle}                                             % Ket.
\newcommand{\ave}[1]{\langle{#1}\rangle}                                       % Statistical average.
\newcommand{\qave}[3]{\langle{#1}|{#2}|{#3}\rangle}                            % Quantum average.
\begin{document}

%%
%%---- Title of the paper --------------------------------------------------------------------------------
%%

\title{Quantum Superposition State Production by Continuous Observations and Feedback}

%%
%%---- Authors and affiliations --------------------------------------------------------------------------
%%

\author{Antonio~Negretti}\email[E-mail: ]{negretti@phys.au.dk}
\author{Uffe~V.~Poulsen}
\author{Klaus M\o lmer}
\affiliation{Lundbeck Foundation Theoretical Center for Quantum System Research\\Department of Physics and
Astronomy, University of Aarhus\\DK-8000 Aarhus C, Denmark}
\date{\today}

%%
%%---- Abstract ------------------------------------------------------------------------------------------
%%

\begin{abstract}
We present a protocol for generation of superpositions of states with
distinguishable field amplitudes in an optical cavity by quantum non-demolition photon number
measurements and coherent feeding of the cavity.
\end{abstract}

%%
%%---- PACS numbers --------------------------------------------------------------------------------------
%%

%% 03.67.-a : Quantum information
%% 02.30.Yy : Control theory
%% 42.50.Dv : Nonclassical states of the electromagnetic field, including entangled photon states;
%%            quantum state engineering and measurements
\maketitle

%%
%%---- Introduction --------------------------------------------------------------------------------------
%%

By suitably tailored optical pulses it is possible to coherently
manipulate the states of small quantum systems and for example to
steer molecular processes and chemical reactions. Methods and concepts
from this research have spread to the field of quantum information
theory which, even with quantum error correction, requires a very high
degree of control \cite{Chen2006}. As an example, quantum
optimal control techniques can substantially improve the performance
of elementary quantum gates with cold neutral atoms
\cite{Treutlein2006}. Optimal control methods aim at manipulating a
few external parameters, e.g., currents and magnetic fields of an
atomic trapping potential, in such a way that an initial state of the
system evolves into the desired final state with high fidelity. These
techniques are open-loop, i.e., they do not exploit the knowledge that
one can get by observing the system and using the measurement outcome
in a suitable feedback. Even quite simple measurements display powers
which are hard to match with controllable interactions in terms of the
states accessible. For example, optical probing of spin-polarized
macroscopic atomic samples has been used to enable atomic
spin-squeezing \cite{Geremia2004}, entanglement,
quantum storage \cite{Julsgaard2004} and teleportation
\cite{Sherson2006}, and measurements of the phase of light transmitted
through a modest cavity has been proposed as a means to project
product states of atoms in the cavity into entangled states and to
implement quantum computation \cite{Sorensen2003}. 

The natural next step is to apply feedback continuously in time using
the information acquired in real time with the measurements. The
theory for continuous measurements and feedback
\cite{Wiseman1994,Belavkin1999} combines the non-deterministic
elements of quantum trajectories \cite{Carmichael1993,Dalibard1992}
with stochastic calculus. While these theories describe correctly the 
outcome of a given measurement and feedback scheme, it is still an 
open problem how one identifies reliable schemes for a given task. A 
scheme for photon Fock state generation in a cavity has been proposed
recently by Geremia in Ref.~\cite{Geremia2006} and an analysis of the
stability of the feedback 
by Yanagisawa~\cite{Yanagisawa2006}. In this Letter we propose
a strategy to generate an equal superposition of two
quantum-mechanical states with distinguishable field amplitudes
$\ket{\Psi} \propto \ket{A} + \,\ket{B}$. Superpositions of states,
which are well localized in separate regions of the effective
position-momentum phase space of the field variables, have been
proposed as useful resources for quantum computation \cite{Knill2001}
and quantum metrology~\cite{Leibfried2005}. 
Other methods for generation of such superposition states have been
demonstrated with light \cite{Raimond1997,Ourjoumtsev2007} and with trapped ions
\cite{Leibfried2005}. Our principal idea is to
apply the method described in Ref.~\cite{Geremia2006} to approach a
target Fock state, occupying a ring in the field amplitude phase
space. Before we reach this state, the phase space distribution is of
crescent shape [see Fig. 1(b)], and we feed coherent radiation into the
cavity to displace the state [Fig. 1(c)]. We subsequently start probing
the photon number again. In a pictorial representation, this scheme
will select a quantum state with phase space support at the overlap of
a new Fock state ring and the displaced crescent distribution, i.e.,
at two crossing regions, and hence a linear superposition of two
quantum states with distinguishable field amplitudes may result from
the protocol [Fig. 1(d)].

A quantum non-demolition measurement of the photon number $\hat{n}$ in
a single cavity mode can be accomplished by measuring the phase shift
of a probe laser field that couples via a cross-Kerr effect to the
cavity field when it passes through an atomic gas inside the cavity.
In our simulations, we apply physical parameters of a dark-state
mechanism in the gas for effective coupling of the fields taken from
Ref.~\cite{Geremia2006}, but the formalism is general and
describes also the effective coupling based on the collective atomic motion
inside the cavity, studied in Ref.~\cite{Gupta2007}. The interaction Hamiltonian is 
proportional to the product of the cavity and the probe photon number 
operators and it causes a phase shift of the probe without exchanging photons 
with the cavity field. It thus enables a non-demolition interaction which
will gradually cause a narrowing of the photon number state
distribution. Our simulations will apply the stochastic master
equation technique~\cite{Wiseman1994,Geremia2006}, but in order to
explain this method and to get useful insight, it is worthwhile to
establish a simple physical picture of the underlying probing
dynamics. For this purpose, consider the probing beam as composed of a 
succession of segments of duration $\Delta t$. The field is in a
coherent state, and hence factors in a product state of coherent
states occupying each segment of the beam. The continuous measurement
on the probe beam after interaction with the cavity field now
separates in the detection on each individual segment of light, and
assuming an incident coherent state with a real field amplitude, the
phase shift is registered by balanced homodyne detection of the
phase-quadrature component by means of interference with a local
oscillator field.

The interaction between the cavity field and a single segment of light
is governed by the unitary operator
$U=e^{-i\,M^{\prime}\,\hat{n}\,\hat{n}_p}$, where $M^{\prime}$ is a
measure of the coupling strength. 
We assume that each segment of the probe beam is in a coherent state 
with a large mean number of photons $\Phi\Delta t$, where $\Phi$ 
is the photon flux. Therefore, we can write  
$\hat{a}_p = \sqrt{\Phi\Delta t} + \hat{a}_p^{\prime}$ and expand the time
evolution operator to lowest order in the quantum fluctuations,
$U=e^{-i\,\beta\,\hat{n}\,\hat{x}_p}$, where
$\beta=2\,M^{\prime}\,\sqrt{\Phi\Delta t}$, and $\hat{x}_p =
(\hat{a}_p^{\prime} + \hat{a}_p^{\prime\,\dagger})/2$. 
The interaction
implies a displacement of the probe $p$ quadrature proportional to
the cavity photon number, and it is thus useful to expand the joint
cavity and probe field state in the corresponding eigenstate basis,
$\ket{\Psi_{\rm in}} = \sum_n c_n |n\rangle \int dp_p \pi^{-1/4}
e^{-p_p^2/2} |p_p\rangle$, so that the state after interaction becomes
\begin{align}
\label{Eqn:outstate} 
\ket{\Psi_{\rm out}} 
= 
\sum_n c_n |n\rangle \int dp_p \pi^{-1/4} e^{-(p_p-\beta n)^2/2} |p_p\rangle.
\end{align}
At this stage, $p_p$ is measured, an arbitrary outcome is obtained
according to the probability distribution,
\begin{align} \label{Eqn:fp}
f(p_p) = \pi^{-1/2}\sum_n|c_n|^2\,e^{-(p_p-\beta\,n )^2},
\end{align}
and the projection of (\ref{Eqn:outstate}) on the corresponding $p_p$
eigenstate, yields the updated state of the cavity field,
\begin{align}
\label{Eqn:condstate}
\begin{split}
  \ket{\Psi_{\rm c}(p_p)} 
  = 
  \mathcal{N}(\beta,p_p)\,\sum_n c_n\,e^{-(\beta\,n - p_p)^2/2}\,\ket{n},
\end{split}
\end{align}
where $\mathcal{N}(\beta,p_p)$ is a normalization factor. When we
model continuous probing, the parameter $\beta$ is infinitesimal, and
the effect of the interaction is merely to shift the Gaussian
probability distribution for $p_p$ by $\beta\langle n\rangle$. The
update of the cavity field state after measurement is also
infinitesimal due to the weak $n$ dependence of the exponential factor
in (\ref{Eqn:condstate}). The random detection can be modelled by a
Wiener noise process, and the conditioned dynamics under continuous
measurements can be brought on the form of an \^Ito stochastic master
equation (SME) \cite{Wiseman1994,Geremia2006}:
\begin{align}
\label{Eqn:SME} 
d \hat\rho(t)
= 
M \, \mathcal{D}[\hat{n}]\hat\rho(t) \, dt + \sqrt{M\,\eta} \,
\mathcal{H} [ \hat{n} ] \hat\rho(t) \, dW(t),
\end{align}
with $\mathcal{D}[\hat{X}]\,\hat\rho \equiv
\hat{X}\,\hat\rho\,\hat{X}^\dagger - 1/2(
\hat{X}^\dagger\,\hat{X}\,\hat\rho + \hat\rho\,
\hat{X}^\dagger\,\hat{X} )$ and $\mathcal{H}[\hat{X}]\,\hat\rho \equiv
\hat{X} \,\hat\rho + \hat\rho\, \hat{X}^\dagger - \mathrm{Tr}[
(\hat{X} + \hat{X}^\dagger)\, \hat\rho ] \,\hat\rho.$ In
(\ref{Eqn:SME}) $M = 2\,M^{\prime\,2}\,\Phi$ denotes the measurement strength and
$\eta\in[0,1]$ represents the quantum efficiency of the detection. The
\emph{ innovation process}, i.e., the difference between the actually
observed $p_p$ and its quantum mechanical expectation value with the
current quantum state of the cavity field, is described by a Wiener
process \cite{Bouten2006}, $dW(t)$. This difference is due to the shot
noise in photo detection. 

While the continuous measurement described above will eventually
collapse the system on one of the Fock states present in the initial
state, an important initial step of our scheme is to evolve the system
towards a \emph{given} Fock state. In Ref.~\cite{Geremia2006}, Geremia
showed how to use the information gradually obtained about $n$ by the
detection record to feed coherent radiation into the cavity which in-
or decreases the total photon number in a controllable manner. This
feedback is described by adding to~(\ref{Eqn:SME}) terms that
describes evolution under the Hamiltonian 
$\hat{H}_\mathrm{Fb}(t) = G\, e_{f}\, \hat{x}$. 
Here $\hat{x}=(\hat a + \hat a^\dagger)/2$ is the cavity field
quadrature operator, $G$ is the feedback gain factor, and $e_{f}$ is the
feedback policy function, that we take to depend on appropriate
expectation values for the field. A natural choice for the feedback
policy function is $e_{f}(\ave{\hat{n}}) = n^{\star} - \ave{\hat{n}}$,
where $n^{\star}$ is the desired photon number \cite{Geremia2006}. The
feedback Hamiltonian causes a displacement of the field quadrature
operator $\hat{p}=(\hat a - \hat a^\dagger)/(2\,i)$. The feedback is
proportional to $n^{\star} - \ave{\hat{n}}$, and a state with
negative $\langle\hat{p}\rangle$ can be shifted to larger negative
values of $\langle\hat{p}\rangle$ and hence typically a larger
$\langle\hat{n}\rangle$, if desired.

\begin{figure}[t]
\begin{center}
\includegraphics{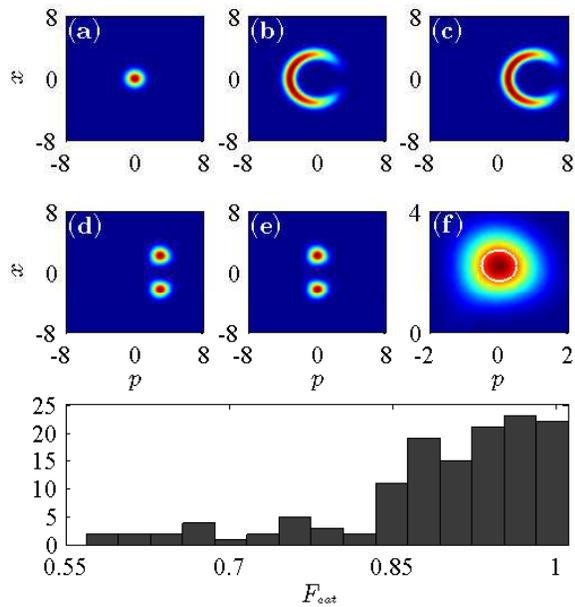}
\end{center}
\caption{(Color online) Time evolution of the Husimi $Q$-function for
  a single successful realization of the linear superposition state
  ($\eta=1$, $\kappa=0$, and $n^{\star}=10$). In (f) the positive $x$
  component of the Husimi $Q$-function in (e) is magnified. The white 
  ring in (f) indicates a contour for the choice of displaced squeezed state
  $\ket{\alpha,\zeta}$ in (\ref{Eqn:CatFidelity}) that yields the best
  overlap with our state. The lower panel shows the histogram of the
  fidelity $\label{Eqn:CatFidelity}$ for 135 simulations with
  $n^{\star}=10$, $\kappa=0$, and $\eta=1$.}
\label{fig:cat}
\end{figure}

The functioning of our complete scheme is illustrated in Fig. 
\ref{fig:cat}(a-e). We find heuristically (for $n^{\star}\simeq 10$)
that the following protocol has a high success rate: At time $t=0$ we
start with the electromagnetic vacuum (a) and we apply the Geremia
Fock state feedback protocol towards $n^{\star}$, but only until the
quadrature $\ave{\hat{p}} < (n^{\star}-1)^{1/4} - \sqrt{n^{\star}}$.
At that time the Husimi $Q$-function looks like a crescent in
position-momentum phase space (b). The state is now shifted towards
the positive value $\ave{\hat{p}}= 0.9\,\sqrt{n^{\star}}$ (c) and at
that time we start again to observe (without feedback) the photon
number in the system until the desired state is found (d). Finally,
the state is shifted to a symmetric state around the centre of the
phase space (e). We stop the action of the probing field when the
maximum value of the Husimi $Q$-function along the $p$ quadrature axis
is below some fixed value $\delta$ (typically $\delta\leq 0.005$).

The protocol does not work in every run of the simulation/experiment.
When we start the measurement on the crescent shape state it sometimes
happens that the state collapses into a quasi coherent state
\cite{Zurek1993}, and sometimes the state dynamics becomes unstable.
Since we have access to the density matrix conditioned on the
measurement outcome, we know if the collapse takes place, and this
problem is partially solved by starting over again to generate a new
crescent state with the number state feedback generator. The
instability problem is solved in our simulations by applying a not too
fast ramp of increased measurement strength $M(t)$, controlled in the
experiment by the probe laser power \cite{Geremia2006}. In our
protocol it is important that the probing laser is not switched on too
fast, but it is also important that it is switched off quickly to not
perturb the state when it has been created ($t_{\rm switch-on} \sim
200$ ns, $t_{\rm switch-off} \sim 1\, -\,2$ ns for $M=2.12$ MHz). 
Using these strategies and allowing a maximal production time of $10\,M^{-1}$, 
we observed a success probability for the production of a quantum 
superposition state of 51\% for $n^{\star}=5$, 69\% for $n^{\star}=8$, 
and 77\% for $n^{\star}=10$. Those results are obtained in the ideal 
situation of perfect detector efficiency and no cavity decay. 

Our protocol does not automatically favor a superposition of coherent
states, and we have investigated to which extent, the state produced
can be written as a superposition of two Gaussian, minimum uncertainty
states in the position-momentum phase space. As a way to quantify the
quality of the state we use the optimal overlap fidelity
\begin{eqnarray}
\label{Eqn:CatFidelity} \mathcal{F}_{\mathrm{sup}} := \max_{\zeta,\alpha,\phi}\left\{
\qave{\Psi(\zeta,\alpha,\phi)}{\hat\rho}{\Psi(\zeta,\alpha,\phi)}\right\},
\end{eqnarray}
where $\ket{\Psi(\zeta,\alpha,\phi)} = \mathcal{N}(\zeta,\alpha,\phi)\,
\left(\ket{\alpha,\zeta} + e^{i\,\phi}\,\ket{-\alpha,\zeta^{\ast}}\right),$ 
and where $\zeta = r e^{i\,\theta}$ is a squeezing parameter, 
$\alpha = x + i\,y$ is a displacement parameter, and 
$\mathcal{N}(\zeta,\alpha,\phi)$ is a normalization factor. The optimal 
superposition state parameters $(\alpha,\zeta,\phi)$ can be determined from 
the detection record in every run of the experiment. The results of 135 
attempts to produce such quantum superposition states are summarized in 
the lower panel of Fig. \ref{fig:cat}, which provide an average fidelity 
of about 90\%. The success probability for $n^{\star}=10$ does not change 
appreciably when a cavity decay rate of $\kappa = 0.005\,M$ and finite detector 
efficiency $\eta=0.8$ are taken into account in our simulations, but 
the average fidelity decreases to 70\%. Note that in our family of test 
superposition states we allow a relative phase $\phi$. This phase is 
known to the experimenter based upon the detection record, but it is not 
under straightforward experimental control.

We will now explain our quantitative findings. We have found
numerically that the crescent shaped state of Fig.~\ref{fig:cat}(b),
which is produced with a high success probability, is very close to
the so-called crescent state of Ref.~\cite{Hradil1991}. These states
are eigenstates of the non-Hermitian operator $\hat{n} - 2\, i
\,|\xi|\, \hat{x}$ \cite{Hradil1991}. 
When $|\xi|\rightarrow 0$ the eigenstate is close to a Fock state, 
while for $|\xi|\gg 1$ it is a coherent state. We now 
insert this state in Eq.~(\ref{Eqn:outstate}) and we can semi-analytically 
follow the effect of probing, without applying any feedback. Since the
measurement is of quantum non-demolition type, the equations (1-3)
discussed for infinitesimal temporal segments, also apply for the
accumulated effect of measuring on the probe field for an extended
period of time.  The parameter $\beta$ is then larger, so that the
probability distribution~(\ref{Eqn:fp}) for the $p_p$ observable for a
long time interval is no longer well approximated by a Gaussian. 
Instead, $\hat{p}_p$ provides a weak measurement of $\hat{n}$, which
according to Eq.~(\ref{Eqn:outstate}) would become a projection in the
limit $\beta\rightarrow\infty$. Figure 2(a) shows a plot of the $p_p$
probability distribution $f(p_p)$ obtained for a typical probing time,
corresponding to $\beta=0.2$, for the crescent state. We identify
three regions on the curves: $(I)$ a maximum for small $p_p$, which is
rather independent of $n^{\star}$, $(II)$ a central region where
$f(p_p)$ decreases, and $(III)$ a second maximum at $p_p\sim
n^{\star}$ followed by a rapid decay. Since $\hat{p}_p$ measures the cavity
photon number $\hat{n}$, the appearance of the three regions can be
qualitatively understood in terms of the $n$ distribution of the state
or, more illustrative, in terms of the Husimi $Q$-function. Fock states are
ring-shaped in phase space and in Fig.~2(c) and 2(d) we plot the
displaced crescent state together with the region borders identified
from Fig.~2(a) for $n^{\star}=8$ and for $n^{\star}=10$. It is seen
that region $(I)$ corresponds to the part of the state close to the
origin. By increasing $n^{\star}$, the weight in this region will only
change slightly.  In region $(II)$ the ``arms'' of the crescent state
are more or less radial in phase space while in region $(III)$ they
come together again. When $n^{\star}$ is increased, the crescent state
becomes larger and the border between region $(II)$ and $(III)$ must
move to larger radii, i.e.\ larger $p_p$.

From the phase space plots it is to be expected that region $(II)$ is
where a weak $n$ measurement will cut out two well separated portions
of the Husimi $Q$-function. This is confirmed by Fig.~2(b), which shows the
maximum value of the post-measurement Husimi $Q$-function on the $p$-axis as
a function of the measured $p_p$. We see that in the central region
$(II)$, the state produced has very small values of the Husimi $Q$-function
on the $p$-axis, while the surrounding regions $(I)$ and $(III)$ are
clearly not useful for our purpose. Since the second maximum in
Fig. 2(a) moves further to the right with increasing $n^{\star}$, the
central region $(II)$ with the successful outcomes becomes larger and
this explains that our protocol works better for larger
$n^{\star}$. 

It should be noticed that the semi-analytical model based on the crescent
state~\cite{Hradil1991} suggests that the optimal superposition state
parameters $(\alpha,\zeta,\phi)$ be uniquely determined by $n^{\star}$,
the duration of the final probing stage, and the corresponding
integrated signal $p_p$. In an experiment, this is much easier than
the numerical optimization based upon the solution of the
SME~(\ref{Eqn:SME}).
\begin{figure}[t]
\begin{center}
\includegraphics[width=8.6cm,height=11.5cm]{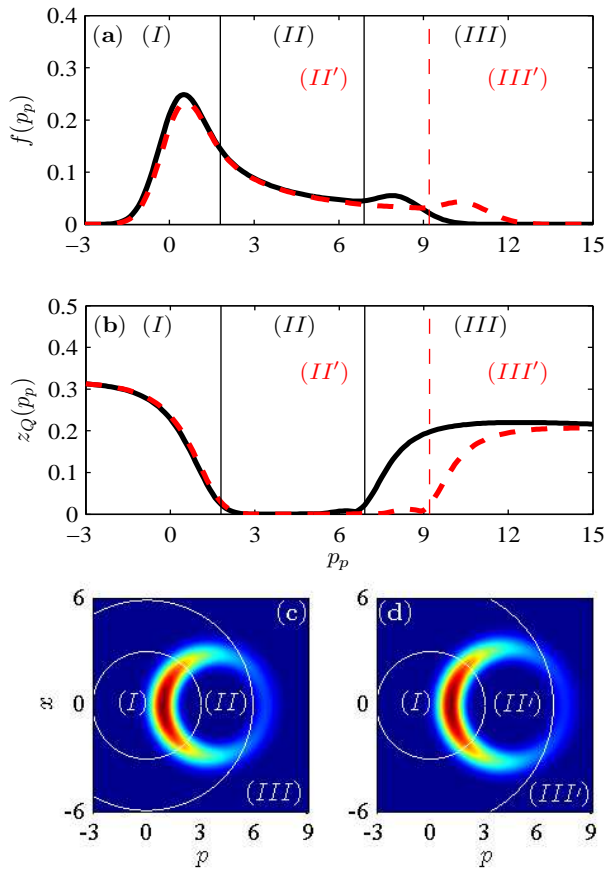}
\end{center}
\caption{(Color online) (a) Probability distribution $f(p_p)$ for the
  outcome of the measurement of the probing field quadrature
  $\hat{p}_p$. (b) Maximum value of the Husimi $Q$-function $Q(x=0,p)$
  plotted as a function of the measurement outcome $p_p$. The black 
  (continuous) lines correspond to $n^{\star}=8$ while the red (dashed) 
  ones to $n^{\star}=10$. In (c) and (d) are shown the corresponding Husimi 
  $Q$-functions of the displaced crescent state together with the region 
  boarders identified from (a) and (b). In all pictures $\beta=0.2$ has 
  been taken.} 
\label{fig:distribution}
\end{figure}

In conclusion we have proposed to generate non-classical states of
light in a cavity by using quantum measurements and feedback.
A protocol for production of highly non-classical superposition states
with high fidelity and success probability was proposed. We are
currently working on the generalization of the ideas presented in this
Letter to the generation of similar states of atomic ensembles.

%%
%%---- Acknowledgments -----------------------------------------------------------------------------------
%%

The authors acknowledge financial support from the European Union
Integrated Project SCALA. A. N.  acknowledges R. J. Hendricks and 
Z. Hradil for useful information, U.V.P. acknowledges financial support by
the Danish Natural Science Research Council, and K.M. acknowledges
support of the ONR MURI on quantum metrology with atomic systems.

%%
%%---- Bibliography --------------------------------------------------------------------------------------
%%

\bibliographystyle{apsrev}
\bibliography{Negrettietal_FinalVersion_LC10997}

\end{document}